\documentstyle [12pt] {article}
\begin{document}

\pagestyle{empty}
\rightline{\vbox{
\halign{&#\hfil\cr
&NUHEP-TH-93-6\cr
&UCD-93-9\cr
&April 1993\cr}}}
\bigskip
\bigskip
\bigskip
{\Large\bf
	\centerline{Perturbative QCD Fragmentation Functions}
	\centerline{for $B_c$ and $B_c^*$ Production}}
\bigskip
\normalsize

\centerline{Eric Braaten and Kingman Cheung}
\centerline{\sl Department of Physics and Astronomy, Northwestern University,
    Evanston, IL 60208}
\bigskip

\centerline{Tzu Chiang Yuan}
\centerline{\sl Davis Institute for High Energy Physics}
\centerline{\sl Department of Physics, University of California,
    Davis, CA  95616}
\bigskip

\begin{abstract}
The dominant production mechanism for ${\bar b} c$ bound states
in high energy processes is the production of a high energy ${\bar b}$
or $c$ quark, followed by its fragmentation into the ${\bar b} c$ state.
We calculate the fragmentation functions for the production
of the S-wave  states $B_c$ and $B_c^*$ to leading order in the QCD
coupling constant.  The fragmentation probabilities for
${\bar b} \rightarrow B_c$ and ${\bar b} \rightarrow B_c^*$
are approximately $2.2 \times 10^{-4}$ and $3.1 \times 10^{-4}$,
while those for $c \rightarrow B_c$ and
$c \rightarrow B_c^*$ are smaller by almost two orders of magnitude.
\end{abstract}

\vfill\eject\pagestyle{plain}\setcounter{page}{1}

The study of the charmonium and bottomonium systems
have played an important role in the development and
eventual acceptance of quantum chromodynamics (QCD) as the field theory of
the strong interactions.  Bound states containing top quarks
may not exist as identifiable states, because the widths of the states due to
top quark decay may be larger than the splittings between the energy levels.
Thus the final frontier for the study of heavy quark-antiquark bound states
may be the ${\bar b} c$ system.
To assess the prospects for the experimental study of ${\bar b} c$ mesons
in present and future colliders, it is important to  have accurate
predictions for their production rates.
The dominant production
mechanism for high energy ${\bar b} c$ mesons is the fragmentation of
${\bar b}$ and $c$ quarks.  The fragmentation process
is described by a universal fragmentation function $D(z)$
that gives the probability for the splitting of the parton into
the desired hadron plus other partons.  For a hadron composed of
only heavy quarks, this fragmentation function
can be calculated using perturbative QCD \cite{by}.
The fragmentation functions for the splitting of gluons and charm quarks
into S-wave charmonium states have been calculated to leading order in
the QCD coupling constant $\alpha_s$ \cite{by,bcy}.
In this paper, we calculate the fragmentation functions for the splitting
of $\bar b$ and $c$ into the $^1S_0$ ground state
$B_c$ and into the $^3S_1$ state $B_c^*$.
These universal fragmentation functions can be used to calculate
the direct production rate of $B_c$ and $B_c^*$ in any high energy
process.

We begin by summarizing previous calculations of the production rates
for ${\bar b} c$ mesons.
A perturbative QCD calculation of the cross section
for $e^+ e^- \rightarrow B_c^* b {\bar c}$ was carried out
by Clavelli in 1982 \cite{cla}.  The cross sections
for $e^+ e^- \rightarrow B_c b {\bar c}$ and
$e^+ e^- \rightarrow B_c^* b {\bar c}$  were subsequently calculated
incorrectly by Amiri and Ji \cite{ja}
and then correctly by Chang and Chen \cite{cc}.  Chang and Chen followed
Amiri and Ji in expressing their results in terms of
fragmentation functions $D(z)$, but they did not realize that these
fragmentation functions are universal and can also be
applied to the hadronic production of $B_c$ and $B_c^*$.
The total production rates for $B_c$ in $e^+ e^-$ and hadron colliders,
including those that are produced by cascade decays from $B_c^*$ and other
$\bar b c$ bound states, have been estimated using a modified version
of the parton shower Monte Carlo HERWIG \cite{lmp}.  The modified version
included two new parameters describing the probabilities for the
production of $c \bar c$ pairs, and the results were sensitive to the values
of the bottom and charm quark masses.
The ratio of the cross section for $B_c$ production and $b$-quark
production was estimated to be
$1.0 \times 10^{-3}$ for $e^+ e^-$ collisions at LEP
and $1.3 \times 10^{-3}$ for $p \bar p$ collisions at the Tevatron,
consistent with a previous crude estimate
\cite{gklst} of the probability for a $\bar b$ to form a
$\bar b c$ meson.  Given the adjustable parameters in the Monte Carlo
calculation of Ref. \cite{lmp},
it is difficult to estimate the errors in its predictions.
It is clear that more reliable calculations of the production rates of
$\bar b c$ mesons would be valuable in assessing the prospects for the
discovery and investigation of these new particles.

The fragmentation contribution to the differential cross section
for the direct production of
a $B_c$ of energy $E$ in any high energy process
is the term that survives in the limit
$E/m_b \rightarrow \infty$.  It has the general form
\begin{equation} {
d\sigma ( B_c(E))
\;=\; \sum_i \int_0^1 dz \;
d{\widehat \sigma}(i(E/z),\mu)
	\; D_{i \rightarrow B_c}(z,\mu) \;,
} \label{fac} \end{equation}
where the sum is over partons of type $i$ and $z$ is the
longitudinal momentum fraction of the $B_c$ relative to the parton.
The physical interpretation of (\ref{fac}) is that a $B_c$
of energy $E$ can be produced by first producing a parton $i$ of larger
energy $E/z$ which subsequently splits into a $B_c$ carrying
a fraction $z$ of the parton energy.
The expression (\ref{fac}) for the differential
cross section has a factored
form:  all the dependence on the energy $E$ is in the parton subprocess
cross section ${\widehat \sigma}$, while all the dependence on the heavy
quark masses $m_c$ and $m_b$
is in the fragmentation function $D_{i \rightarrow B_c}$.
The dependence on the arbitrary factorization
scale $\mu$ cancels between the two factors.
Large logarithms of $E/\mu$ in the subprocess cross section
${\widehat \sigma}$ can be avoided by choosing $\mu$ on the order of $E$.
Large logarithms of $\mu/m_b$ then necessarily appear in the
fragmentation functions $D_{i \rightarrow B_c}(z,\mu)$,
but they can be summed up by solving the evolution equations \cite{rdf}
\begin{equation} {
\mu {\partial \ \over \partial \mu} D_{i \rightarrow B_c}(z,\mu)
\;=\; \sum_j \int_z^1 {dy \over y} \; P_{i\rightarrow j}(z/y,\mu)
	\; D_{j \rightarrow B_c}(y,\mu) \;,
} \label{evol} \end{equation}
where $P_{i\rightarrow j}(x,\mu)$ is the Altarelli-Parisi
function for the splitting of the parton of type $i$ into a parton of
type $j$ with longitudinal momentum fraction $x$.
For example, the ${\bar b} \rightarrow {\bar b}$ splitting function for
a bottom antiquark with energy much greater than its mass is the usual
splitting function for quarks and antiquarks:
\begin{equation} {
P_{{\bar b} \rightarrow {\bar b}}(x,\mu)
\;=\; {2 \alpha_s(\mu) \over 3 \pi}
	\left( {1 + x^2 \over 1-x} \right)_+ \;,
} \label{Pc} \end{equation}
where $f(x)_+ = f(x) - \delta(1-x) \int_0^1 f(x') dx'$.
The boundary condition on the evolution equation
(\ref{evol}) is the initial fragmentation function
$D_{i \rightarrow B_c}(z,\mu_0)$ at some scale $\mu_0$ of order $m_b$.
The initial fragmentation function
can be calculated perturbatively as a series in $\alpha_s(\mu_0)$.
At leading order in $\alpha_s$, the initial
fragmentation function for $b \rightarrow B_c$
can be calculated using the expression
\cite{bcy2}
\begin{equation}{
D_{{\bar b} \rightarrow B_c}(z)
\;=\; {1 \over 16 \pi^2} \int ds
\; \theta \left( s - {(m_b + m_c)^2 \over z} - {m_c^2 \over 1-z} \right)
\lim_{q_0 \rightarrow \infty} {|{\cal M}|^2 \over |{\cal M}_0|^2} \;,
} \label{Dfrag} \end{equation}
where ${\cal M}$ is the matrix element for the production of a $B_c$
and $\bar c$ with total 4-momentum $q$ and invariant mass $s = q^2$
and ${\cal M}_0$ is the matrix
element for the production of a $\bar b$ with the same 3-momentum $\vec q$.
If the momentum of the $B_c$ is $p = (p_0,p_1,p_2,p_3)$ in a frame where
$q = (q_0,0,0,q_3)$, then the longitudinal momentum fraction $z$
is defined by $z = (p_0+p_3)/(q_0+q_3)$.
The limit $q_0 \rightarrow \infty$ in (\ref{Dfrag})
can be taken in any Lorentz frame where $q_3/q_0 \rightarrow 1$.

We proceed to calculate the matrix element ${\cal M}$
for the production of $B_c$ in the fragmentation limit $q_0 >> m_b$.
If $q = (q_0,0,0,q_3)$, it is convenient to carry out the calculation
in the axial gauge associated with the 4-vector
$n = (1,0,0,-1)$.  In this gauge, the fragmentation contribution
comes only from the Feynman diagram for
${\bar b}^* \rightarrow \bar b c \bar c$ shown in Figure 1.
In covariant gauges, there are also contributions from other diagrams
but Ward identities can be used to manipulate the sum of the diagrams into
an expression equivalent to that obtained directly from Figure 1 in axial
gauge.
The amplitude for Figure 1 in this gauge can be reduced to
\begin{eqnarray}
{\cal M} \;=\;  {g_s^2 R(0) \over 3 \sqrt{3 \pi M}}
{1 \over r (s - m_b^2)^2} \;
{\bar \Gamma} \Bigg( 2 \; ({\not \! q} + m_b) \; ({\not \! p} - 2 M)
	 \; \gamma_5
\nonumber \\
\;-\; {s - m_b^2 \over n \cdot (q - (1-r) p)} \;
	({\not \! p} + M) \; \gamma_5 \; {\not \! n}
\Bigg) \; v(p')  \;,
\label{MBc} \end{eqnarray}
where $r = m_c/(m_b+m_c)$, $M = m_b+m_c$, and $R(0)$ is the nonrelativistic
radial wavefunction at the origin for the $B_c$.
The Dirac spinor $\bar \Gamma$ is the matrix element for the production
of a ${\bar b}$ of momentum $q = p + p'$. Its explicit form is not needed
to calculate the fragmentation function (\ref{Dfrag}).
Summing over final spins and colors, the square of the
matrix element (\ref{MBc}) reduces to
\begin{equation} {
|{\cal M}|^2 \;=\; {64 \pi \alpha_s^2 |R(0)|^2 \over 27 M} \;
{1 \over r^2 (s - m_b^2)^4}
{\rm tr} \Bigg( \Gamma \; {\bar \Gamma} \; \Delta \Bigg) \;,
} \label{Msq} \end{equation}
where $\Delta$ is a Dirac matrix that depends on $q$ and $p$.
We need only keep the terms in $\Delta$ that are of order $m_b^4 q_0$.
While $q$ and $p$ both have components of order $q_0$, the invariant mass
$s = q^2$ is of order $m_b^2$ in the fragmentation region.
Simplifying the Dirac matrix $\Delta$ by dropping terms which
are suppressed by powers of $m_b/q_0$, it reduces to
\begin{eqnarray}
\Delta &=&
\left( s^2 - 2(2-4r+r^2) M^2 s + (1-r)(3-r)(1-4r+r^2) M^4 \right)
	\; {\not \! q}
\nonumber \\
&-& \left( s - (2-r)^2 M^2 \right) \left( s - (1-r)^2 M^2 \right) \; {\not \!
p}
\nonumber \\
&+& 2 \; {n \cdot p  \over n \cdot q - (1-r) n \cdot p}
	\; (s - (1-r)^2 M^2)^2 \; ({\not \! q} - {\not \! p})
\nonumber \\
&-& 2 \; {n \cdot p \over n \cdot q - (1-r) n \cdot p} M^2 (s - (1-r)^2 M^2)
	\left( (1-2r)\; {\not \! q} \;-\; (1-r) \; {\not \! p} \right)
\nonumber \\
&+& {n \cdot p \; n \cdot (q-p) \over (n \cdot q - (1-r) n \cdot p)^2}
		(s - (1-r)^2 M^2)^2 \; {\not \! p} \;.
\label{DelBc} \end{eqnarray}
Since the coefficients of ${\not \! p}$ and ${\not \! q}$
in (\ref{DelBc}) are all manifestly of order $m_b^4$, we can make the
substitution $p = zq$ which is accurate to leading order in $m_b/q_0$.
The Dirac trace in (\ref{DelBc}) is then proportional to
${\rm tr}(\Gamma {\bar \Gamma} {\not \! q})$.
The square of the matrix element for the production of a real ${\bar b}$
of momentum $q$ is
$|{\cal M}_0|^2 = {\rm tr}(\Gamma {\bar \Gamma} ({\not \! q} - m_b))$,
which reduces in the fragmentation limit to
${\rm tr}(\Gamma {\bar \Gamma} {\not \! q})$.
Dividing $|{\cal M}|^2$ in (\ref{Msq}) by $|{\cal M}_0|^2$
and inserting it into (\ref{Dfrag}), we obtain the fragmentation function
as an integral over $s$:
\begin{eqnarray}
D_{{\bar b} \rightarrow B_c}(z)
\;=\; {4 \alpha_s^2 |R(0)|^2 \over 27 \pi m_c^3}
\int ds \; \theta \left(s - {M^2 \over z} - {m_c^2 \over 1-z} \right)
\Bigg( { (1-z)(1+rz)^2 r M^2 \over (1-(1-r)z)^2 (s - m_b^2)^2}
\nonumber \\
\;-\; {( 2(1-2r) - (3-4r+4r^2)z + (1-r)(1-2r)z^2 ) r M^4
	\over (1-(1-r)z) (s - m_b^2)^3}
\;-\; {4 r^2 (1-r) M^6 \over (s - m_b^2)^4} \Bigg) \;.
\label{DBcint} \end{eqnarray}
Evaluating the integral over $s$,
we obtain our final expression for the fragmentation function
at the initial scale $\mu_0$:
\begin{eqnarray}
D_{{\bar b} \rightarrow B_c}(z,\mu_0)
&=& {2 \alpha_s(\mu_0)^2 |R(0)|^2 \over 81 \pi m_c^3}
\; {r z (1-z)^2 \over (1 - (1-r)z)^6}
\Bigg(6 \;-\; 18(1-2r)z \;+\; (21-74r+68r^2)z^2
\nonumber \\
&-& 2(1-r)(6-19r+18r^2)z^3 \;+\; 3(1-r)^2(1-2r+2r^2)z^4 \Bigg) \;,
\label{DBc} \end{eqnarray}
where $r = m_c/(m_b+m_c)$.  This agrees with the result of
Chang and Chen \cite{cc} up to the overall normalization.
Chang and Chen normalized their fragmentation functions so that
they integrated to 1: $\int_0^1 dz D(z) = 1$.
Setting $m_b=m_c$ in (\ref{DBc}), we also recover the fragmentation function
for $c \rightarrow \eta_c$ calculated in Ref. \cite{bcy}.
The incorrect result of Amiri and Ji \cite{ja} can be obtained by
keeping only the term proportional to $1/(s-m_b^2)^2$
in the integrand of (\ref{DBcint}).

The initial scale $\mu_0$ in the
fragmentation function and in the running coupling constant in (\ref{DBc})
should be chosen on the order of $m_b$ and $m_c$.
We would like a more specific
prescription for $\mu_0$ that remains physically sensible
even when $m_c/m_b$ is much less than or much greater than 1.
In the Feynman diagram in Figure 1, the invariant mass of the gluon is
$r (s - m_b^2)$ and the virtual $\bar b$ is off shell by
$s - m_b^2$.  For generic values of $z$, $s$
is of order $m_b^2$, and the $\bar b$ and the gluon are off-shell by
amounts of order $(m_b+m_c)^2$ and $m_c (m_b+m_c)$, respectively.
However a large part of the fragmentation probability
comes from the region where $s$ is near its minimum value $(m_b+2 m_c)^2$
and the virtual $\bar b$ and the gluon are only
off-shell by $4 m_c (m_b+m_c)$ and $4 m_c^2$.  As a compromise
between the various scales in the problem, we take
$\mu_0 = \sqrt{4 m_c (m_b+m_c)}$.
In Ref. \cite{cc}, Chang and Chen
used the value $\alpha_s = 0.15$ for the running coupling constant,
which corresponds to a scale $\mu_0$ of order $M_Z$.  As shown by our analysis,
the scale for the fragmentation process is definitely set by the heavy
quark masses.  Thus Chang and Chen underestimated the rates for
production of $B_c$ and $B_c^*$ in $Z^0$ decay by about a factor of 2.

The fragmentation function for a ${\bar b}$ to split into the $^3S_1$
state $B_c^*$ can be calculated in the same way as for $B_c$.
The matrix element for the production of $B_c^* {\bar c}$ is
\begin{eqnarray}
{\cal M} \;=\;  {g_s^2 R(0) \over 3 \sqrt{3 \pi M}}
\epsilon_\mu(p)^* {1 \over r (s - m_b^2)^2}
\; {\bar \Gamma} \Bigg( 2 M \; ({\not \! q} +  m_b) \; \gamma^\mu
\nonumber \\
\;+\; {s - m_b^2 \over n \cdot (q - (1-r) p)} \;
	 ({\not \! p} + M) \; \gamma^\mu \; {\not \! n} \Bigg) \; v(p') \;,
\label{MBstar} \end{eqnarray}
where $\epsilon_\mu(p)$ is the polarization vector for $B_c^*$.
Following the same path as in the $B_c$ calculation, we obtain an
expression for the fragmentation function as an integral over the
invariant mass $s$:
\begin{eqnarray}
D_{{\bar b} \rightarrow B_c^*}(z)
\;=\; {4 \alpha_s^2 |R(0)|^2 \over 27 \pi m_c^3}
\int ds \; \theta \left(s - {M^2 \over z} - {m_c^2 \over 1-z} \right)
\Bigg( { (1-z)(1+2rz+(2+r^2)z^2) r M^2 \over (1-(1-r)z)^2 (s - m_b^2)^2}
\nonumber \\
\;-\; {( 2(1+2r) - (1+12r-4r^2)z - (1-r)(1+2r)z^2 ) r M^4
	\over (1-(1-r)z) (s - m_b^2)^3}
\;-\; {12 r^2 (1-r) M^6 \over (s - m_b^2)^4} \Bigg) \;.
\label{DBstarint} \end{eqnarray}
Integrating over $s$,
the final result for the fragmentation function
for $\bar b$ to split into $B_c^*$ is
\begin{eqnarray}
D_{{\bar b} \rightarrow B_c^*}(z,\mu_0)
&=& {2 \alpha_s(\mu_0)^2 |R(0)|^2 \over 27 \pi m_c^3} \;
{r z (1-z)^2 \over (1 - (1-r) z)^6}
\Bigg(2 \;-\; 2(3-2r)z \;+\; 3(3-2r+4r^2)z^2
\nonumber \\
&-& 2(1-r)(4-r+2r^2)z^3
	\;+\; (1-r)^2(3-2r+2r^2)z^4 \Bigg) \;.
\label{DBstar} \end{eqnarray}
This agrees with the result of Chang and Chen \cite{cc} up to the
overall normalization.  Setting $m_b = m_c$ in (\ref{DBstar}),
we recover the fragmentation function for
$c \rightarrow \psi$ calculated in Ref. \cite{bcy}.
The incorrect result of Amiri and Ji \cite{ja} can be obtained
by keeping only the term proportional to $1/(s-m_b^2)^2$
in the integrand of (\ref{DBstarint}).

The fragmentation functions for a charm quark to split
into $B_c$ and $B_c^*$ can be obtained from (\ref{DBc}) and (\ref{DBstar})
by interchanging the masses $m_b$ and $m_c$:
\begin{eqnarray}
D_{c \rightarrow B_c}(z,\mu_0')
&=& {2 \alpha_s(\mu_0')^2 |R(0)|^2 \over 81 \pi m_b^3} \;
{(1-r) z (1-z)^2 \over (1 - rz)^6}
\Bigg(6 \;+\; 18(1-2r)z \;+\; (15-62r+68r^2)z^2
\nonumber \\
&-& 2r(5-17r+18r^2)z^3 \;+\; 3r^2(1-2r+2r^2)z^4 \Bigg) \;,
\label{cDBc} \end{eqnarray}
\begin{eqnarray}
D_{c \rightarrow B_c^*}(z,\mu_0')
&=& {2 \alpha_s(\mu_0')^2 |R(0)|^2 \over 27 \pi m_b^3} \;
{(1-r) z (1-z)^2 \over (1 - rz)^6}
\Bigg(2 \;-\; 2(1+2r)z \;+\; 3(5-6r+4r^2)z^2
\nonumber \\
&-& 2r(5-3r+2r^2)z^3 \;+\; r^2(3-2r+2r^2)z^4 \Bigg) \;,
\label{cDBstar} \end{eqnarray}
where $r = m_c/(m_b+m_c)$ and $\mu_0' = \sqrt{4m_b(m_b+m_c)}$.

The fragmentation functions for $\bar b \rightarrow B_c$ and
$\bar b \rightarrow B_c^*$ are shown in Figure 2.   The radial
wavefunction at the origin for the $B_c$ and $B_c^*$ has been
estimated from potential models to be $|R(0)|^2 = (1.18 \; {\rm GeV})^3$
\cite{eq}.  For the quark masses,
we use $m_b = 4.9 \; {\rm GeV}$ and $m_c = 1.5 \; {\rm GeV}$.
The solid curves in Figure 2 are the initial fragmentation
functions at the scale $\mu = 6.2 \; {\rm GeV}$, where the running coupling
constant has the value $\alpha_s(\mu) = 0.20$.  The dotted curves
are the fragmentation functions with the scale $\mu$
evolved up to $62 \; {\rm GeV}$ using
the Altarelli-Parisi equation (\ref{evol}).
The average value of the momentum fraction
for $\bar b \rightarrow B_c$ is $<\!\!z\!\!> = 0.68$ at
$\mu = 6.2  \: {\rm GeV}$, and it shifts
downward to $<\!\!z\!\!> = 0.61$
when the scale is evolved up by a factor of 10.
The process $\bar b \rightarrow B_c^*$ has a slightly harder distribution
with $<\!\!z\!\!>= 0.73$ at $\mu = 6.2  \: {\rm GeV}$
and $<\!\!z\!\!> = 0.66$ at $\mu = 62  \: {\rm GeV}$.
The fragmentation functions for $c \rightarrow B_c$ and
$c \rightarrow B_c^*$ are shown in Figure 3.
The solid curves are the initial fragmentation
functions at the scale $\mu = 11.2 \; {\rm GeV}$, where the running coupling
constant has the value $\alpha_s(\mu) = 0.17$.  The fragmentation
functions at the scale $\mu = 112 \; {\rm GeV}$ are shown as dotted
curves in Figure 3.
The average value of the momentum fraction $z$ is initially
0.51 for $c \rightarrow B_c$ and 0.55 for $c \rightarrow B_c^*$.
When the scale is evolved up by a factor of 10, they shift downward
by about $9 \%$.  The evolution of the fragmentation functions would shift
the momentum distributions for the $B_c$ and $B_c^*$
produced in $Z^0$ decay calculated by Chang and Chen \cite{cc}
to slightly smaller values of $z$.

The evolution equation (\ref{evol}) implies that at leading order
in $\alpha_s$, the fragmentation probability $\int_0^1 dz D(z,\mu)$
does not evolve with the scale $\mu$.  The evolution merely shifts the
$z$ distribution to smaller values of $z$.
Therefore the fragmentation probabilities can be used
to provide simple estimates of production rates.  We need only multiply
the production rates for bottom and charm quarks with energies much larger
than their masses by the appropriate fragmentation probabilities.
The fragmentation probabilities for the production of the $B_c$ are
\begin{equation} {
\int_0^1 dz \; D_{{\bar b} \rightarrow B_c}(z,\mu_0) \;=\;
{2 \alpha_s(\mu_0)^2 |R(0)|^2 \over 27 \pi m_c^3}
	\; f \left({m_c \over m_b + m_c}\right) \;,
} \label{PBc} \end{equation}
\begin{equation} {
\int_0^1 dz \; D_{c \rightarrow B_c}(z,\mu_0') \;=\;
{2 \alpha_s(\mu_0')^2 |R(0)|^2 \over 27 \pi m_b^3}
	\; f \left({m_b \over m_b+m_c}\right) \;,
} \label{cBc} \end{equation}
where the function $f(r)$ is
\begin{equation} {
f(r) \;=\; {8 + 13r + 228r^2 - 212r^3 + 53r^4 \over 15 (1-r)^5}
\;+\; {r (1 + 8r + r^2 - 6r^3 + 2 r^4) \over (1-r)^6} \log(r) \;.
} \label{fr} \end{equation}
This function is almost constant,  varying only from $0.496$ at $r=0.23$
to $0.464$ at $r=0.77$.
The fragmentation probabilities for the production of the $B_c^*$ are
\begin{equation} {
\int_0^1 dz \; D_{{\bar b} \rightarrow B_c^*}(z,\mu_0) \;=\;
{2 \alpha_s(\mu_0)^2 |R(0)|^2 \over 27 \pi m_c^3}
	\; g \left({m_c \over m_b+m_c}\right) \;,
} \label{PBstar} \end{equation}
\begin{equation} {
\int_0^1 dz \; D_{c \rightarrow B_c^*}(z,\mu_0') \;=\;
{2 \alpha_s(\mu_0')^2 |R(0)|^2 \over 27 \pi m_b^3}
	\; g \left({m_b \over m_b+m_c}\right) \;,
} \label{cBstar} \end{equation}
where the function $g(r)$ is
\begin{equation} {
g(r) \;=\;
{24 + 109 r - 126 r^2 + 174 r^3 + 89 r^4 \over 15 (1-r)^5} \;+\;
{r (7 - 4 r + 3 r^2 + 10 r^3 + 2 r^4) \over  (1-r)^6} \log(r) \;.
} \label{gr} \end{equation}
It varies from $0.702$ at $r=0.23$ to $0.400$ at $r=0.77$.
The fragmentation probability (\ref{PBstar}) agrees with the
result for $e^+ e^- \rightarrow B_c^* b {\bar c}$ \cite{cla}
obtained by Clavelli.  The numerical values for
the fragmentation probabilities
are $2.2 \times 10^{-4}$ for $\bar b \rightarrow B_c$,
$3.1 \times 10^{-4}$ for $\bar b \rightarrow B_c^*$,
$4.6 \times 10^{-6}$ for $c \rightarrow B_c$,
and $3.9 \times 10^{-6}$ for $c \rightarrow B_c^*$.
The fragmentation probabilities for $c$
are smaller than those for $\bar b$ by almost
two orders of magnitude, because they are proportional to $1/m_b^3$
instead of $1/m_c^3$.

It is interesting to compare our perturbative QCD fragmentation functions
with the phenomenological Peterson fragmentation function \cite{pssz}
which is often used to describe the fragmentation of a $b$ quark into
a hadron. For the case of a $\bar b c$ hadron $H$,
it is proportional to the energy denominator
for the transition of a high momentum $\bar b$ into a $\bar b c$
pair with zero relative velocity and a $\bar c$:
\begin{equation} {
D_{b \rightarrow H}(z) \;=\;
N_H {z (1-z)^2 \over (1-z)^2 + \epsilon z} \;,
} \label{DP} \end{equation}
where $\epsilon = m_c^2/m_b^2$ and $N_H$ is a phenomenological parameter.
The Peterson fragmentation function has the same qualitative behavior
as the functions (\ref{DBc}) and (\ref{DBstar}) predicted by perturbative QCD.
The main difference is that the QCD calculation predicts the normalizations
of the fragmentation functions, while they are given by
arbitrary phenomenological parameters $N_H$ in the Peterson formula.
The Peterson formula (\ref{DP}) also has the same shape for all
$\bar b c$ hadrons $H$, while the precise shape of the QCD fragmentation
functions depends on the orbital and spin quantum numbers.

The only inputs into our perturbative QCD calculations of the fragmentation
functions are the QCD coupling constant $\alpha_s$,
the quark masses $m_b$ and $m_c$, and
the nonrelativistic radial wavefunction at the origin $R(0)$.
The potential model values for $R(0)$ should be quite reliable,
since these models are tuned to reproduce the spectra of the charmonium
and bottomonium systems
and the $\bar b c$ system represents an intermediate case.
Potential models should also give accurate results for ratios of quark masses
such as $r = m_c/(m_b + m_c)$.   The largest uncertainty in the fragmentation
functions for $\bar b \rightarrow B_c, B_c^*$ should therefore come from the
overall factor of $1/m_c^3$. Allowing for a variation of $0.2 \; {\rm GeV}$
in the value $m_c = 1.5 \; {\rm GeV}$, the uncertainty in the normalization
is an overall multiplicative factor of 1.5.
Similarly, the largest uncertainty in the fragmentation
functions for $c \rightarrow B_c, B_c^*$ comes from the
overall factor of $1/m_b^3$.  Allowing for a variation of $0.3 \; {\rm GeV}$
in the value $m_b = 4.9 \; {\rm GeV}$, the uncertainty is an overall
multiplicative factor of 1.2 in the normalization.

The fragmentation functions $D_{\bar b \rightarrow B_c}(z,\mu)$ and
$D_{c \rightarrow B_c}(z,\mu)$ can be used to calculate
the rate for the direct production of $B_c$ in any high energy process.
To obtain the total production rate for $B_c$, we also need the direct
production rates for all the other $\bar b c$ states below the $B D$
threshold, since they all decay ultimately into $B_c$ with branching
fractions near $100 \%$.  We have calculated the fragmentation functions
for the production of the $^3S_1$ state $B_c^*$.  Our fragmentation
functions can also be applied directly to the production of the first radial
excitations of the $B_c$ and $B_c^*$ simply by changing the wavefunction
at the origin.  From potential model calculations \cite{eq}, the value
for the first radial excitation of the S-wave states is
$|R(0)|^2 = (0.99 \; {\rm GeV})^3$.
Thus the production rates for these states
should be about $60 \%$ of those for the $B_c$ and $B_c^*$.
The second radial excitation is probably above
the $B D$ threshold, so its branching fraction into $B_c$ will be negligible.
Adding up all the S-wave contributions to the $B_c$ production rate,
the total fragmentation probabilities
are about $8.5 \times 10^{-4}$ from the splitting of the $\bar b$ and
$1.4 \times 10^{-5}$ from the splitting of the charm quark.
Our lower bound of $8.5 \times 10^{-4}$ for the probability for a high
energy $b$-quark to form a $\bar b c$ bound state is consistent
with the Monte Carlo estimates obtained in Ref. \cite{lmp}.

An accurate calculation of the total production
rate for $B_c$ could probably be obtained by including also
the two sets of P-wave ${\bar b} c$ states below $B D$ threshold.
In calculating the production of the
P-wave states, there are two distinct contributions that must
be included at leading order in $\alpha_s$.
The P-wave state can arise either from the
production  of a collinear $\bar b c$ pair in a color-singlet P-wave state,
or from the production of a collinear $\bar b c$ pair in a color-octet
S-wave state \cite{bbly}.  There are two nonperturbative parameters that enter
into the calculation.  In addition to the derivative of the radial wavefunction
at the origin for the P-wave state, we also require
the probability density at the origin for the ${\bar b}$
and $c$ to be in a color octet S-wave state accompanied by a soft gluon.

Our results should be useful for assessing the prospects for the discovery
of the $B_c$ in $Z^0$ decay at LEP or in $p \bar p$ collisions at the
Tevatron.  The branching fraction for $Z^0 \rightarrow b \bar b$ is
approximately $15.2 \%$.  Multiplying by the total fragmentation
probability for the production of S-wave $\bar b c$ states, we obtain
a lower bound on the inclusive branching fraction for the production of
$B_c$ of $1.3 \times 10^{-4}$.
The contribution from $Z^0 \rightarrow c \bar c$ followed by fragmentation
of the charm quark or antiquark is smaller by about a factor of 80.
At the Tevatron, the cross section for the production of charm quarks at large
$p_T$ is a little larger than for bottom quarks, but not enough  to make
up for the difference in the fragmentation probabilities.  Thus the dominant
production mechanism for $\bar b c$ states will again be $\bar b$
fragmentation.
The cross section for the inclusive production of $B_c$ with
$p_T > 20 \; {\rm GeV}$ at the Tevatron
can be estimated by multiplying the cross section for
the production of $b$ quarks with $p_T > 20 \; {\rm GeV}$, which is measured
to be about 400 nb \cite{cdf}, by the fragmentation probability
of $8.5 \times 10^{-4}$.  Simple estimates of the production rates
at future colliders can be obtained in the same way.

We have pointed out in this paper that the dominant mechanism
for the direct production of high energy $\bar b c$ mesons
is fragmentation, the production of a high energy $\bar b$
or $c$ quark followed by its splitting into the the $\bar b c$ state.
We calculated the fragmentation functions $D(z)$ for production of the S-wave
states $B_c$ and $B_c^*$ to leading order in $\alpha_s$.
These fragmentation functions are universal, so they can be used to
calculate the rates and momentum distributions for the
direct production of $B_c$ and $B_c^*$ in any high energy process.

This work was supported in part by the U.S. Department of Energy,
Division of High Energy Physics, under Grant DE-FG02-91-ER40684.

\vfill \eject

\bigskip
\noindent{\Large\bf Figure Captions}
\begin{enumerate}
\item Feynman diagram for $\bar b^* \rightarrow \bar b c \bar c$
which contributes to the fragmentation of $\bar b$ into $B_c$ and $B_c^*$.
\item Fragmentation functions $D_{{\bar b} \rightarrow B_c}(z,\mu)$
and $D_{{\bar b} \rightarrow B_c^*}(z,\mu)$ as a function of $z$
at $\mu = 6.2 \; {\rm GeV}$ (solid line)
and $\mu = 62 \; {\rm GeV}$ (dotted line).
\item Fragmentation functions $D_{c \rightarrow B_c}(z,\mu)$
and $D_{c \rightarrow B_c^*}(z,\mu)$ as a function of $z$
at $\mu = 11.2 \; {\rm GeV}$ (solid line)
and $\mu = 112 \; {\rm GeV}$ (dotted line).
\end{enumerate}
\vfill\eject

\end{document}